\documentclass[aps,prb,showpacs,reprint,superscriptaddress]{revtex4-1}

\usepackage{graphicx}
\usepackage{amsmath}
\usepackage{bbm}
\usepackage{color}

\begin{document}
\title
{Tunneling conductance in half-metal/conical magnet/superconductor
junctions in the adiabatic and non-adiabatic regime: self-consistent calculations}
\author{P. W\'ojcik}
\email[Electronic address: ]{pawel.wojcik@fis.agh.edu.pl}
\author{B. Rzeszotarski}
\affiliation{AGH University of Science and Technology, Faculty of Physics and Applied
Computer Science, al. Mickiewicza 30, Krak\'ow, Poland}
\author{M. Zegrodnik}
\affiliation
{AGH University of Science and Technology, Academic Center for Materials and Nanotechnology, al. A.
Mickiewicza 30, Krak\'ow, Poland}

\begin{abstract} 
The tunneling conductance in the half-metal/conical magnet/superconductor (HM/CM/SC) is
investigated by the use of the combined Blonder-Tinkham-Klapwijk (BTK) formalism and the
Bogoliubov-de Gennes (BdG) equations. We show that the 
conductance calculated self-consistently differs significantly from the one calculated in the
non-self-consistent framework. The use of the self-consistent procedure ensures that the charge
conservation is satisfied. Due to the spin band separation in the HM, the conductance in the
subgap region is mainly determined by the anomalous Andreev reflection the probability of which
strongly depends on the spin transmission in the CM layer. We show that the spin of electron  
injected from the HM can be transmitted through the CM to the SC adiabatically or non-adiabatically
depending on the period of the exchange field modulation. We find that the conductance in the subgap
region oscillates as a function of the CM layer thickness wherein the oscillations transform from
irregular, in the non-adiabatic regime, to regular in the adiabatic case. In the non-adiabatic
regime the decrease of the exchange field amplitude in the CM leads to the emergence of the
conductance peak for one particular CM thickness in agreement with experiment [J.W.A Robinson, J. D.
S Witt and M. G. Blamire, Science 329, 5987]. For both transport regimes the conductance is
analyzed over a broad range of parameters determining the spiral magnetization in the CM.
\end{abstract}

\maketitle
\section{Introduction}
\label{sec:intro}
In recent years the quantum transport in the ferromagnet/superconductor (FM/SC) junctions has
attracted growing interest due to a possible existence of the spin-triplet
pairing\cite{Champel2005,Braude2007,Houzet2007,Lofwander2005} and  novel
transport phenomena related to this unique superconducting state.\cite{Buzdin2005,Bergeret2005} 
For the s-wave superconductors with the spatially symmetric orbital part of the Cooper pair wave
function, the Pauli principle requires that its spin part  is antisymmetric, which means that
the spin-singlet seems to be the only possible state for the Cooper pair. However, many years ago
Berezinskii\cite{Berezinskii1974} proposed a possible existence of the spin-triplet
state in a system with the s-wave interaction which do not violated the Pauli principle. The triplet
pairing correlations proposed by Berezinskii are odd in time (or frequency) and can appear 
in systems with sort of the time-reversal symmetry breaking mechanism. Recent studies
suggest that the spin-triplet Cooper pair correlations can be induced and observed
experimentally in the FM/SC junctions with the spin-active or magnetically inhomogeneous
interface.\cite{Bergeret2001,Bergeret2003,Volkov2003}

In the normal metal/superconductor (NM/SC) junctions the electrons incident on the interface from
the NM side are reflected as holes with opposite spins. This mechanism, known as the Andreev
reflection,\cite{Andreev1964} leads to the proximity effect - the superconducting pairing
correlations penetrate into the normal metal over the distance as long as one micron at low
temperature.\cite{Lambert1998} 
The proximity effect significantly changes if we replace the normal metal by the ferromagnet. The
exchange interaction in the ferromagnet results in the different Fermi wave vectors for
electrons with opposite spins forming the Cooper pairs. This wave vector mismatch is compensated by
the non-zero total momentum of the electron pairs giving rise to the oscillations of the
spin-singlet superconducting correlations in the ferromagnet,\cite{Buzdin2005,Demler1997} known as
the FFLO oscillations.\cite{Fulde1964,Larkin1964} Since the exchange field tends to align the
electronic spins along the field direction, the spin-singlet superconducting
correlations in the ferromagnet are strongly suppressed leading to the short-range penetration
length. In contrast to the short-range proximity effect for the spin-singlet state, the spin-triplet
state with $m=\pm 1$, with both electronic spin of the Cooper pair directed along the
exchange field, is robust against the pair breaking induced by the exchange interaction. Therefore,
the spin-triplet superconducting correlations ($m=\pm 1$), if they exist, can penetrate the
ferromagnet over the distance comparable to this observed in the NM/SC junctions. This phenomenon,
called long-range proximity effect was predicted theoretically by Bergeret et al.
(see Refs.~\onlinecite{Bergeret2001,Bergeret2003,Bergeret2005}). It was
found\cite{Bergeret2001,Bergeret2003,Bergeret2005} that in the FM/SC multilayer junctions with the
spin-active or magnetically inhomogeneous interface (the spin-flip processes are possible) all three
components $m=0$ and $m=\pm 1$ of the spin-triplet state can arise. Despite few theoretical
studies on the spin-triplet pairing induced by the spin-active interface, including the effect of
domain wall,\cite{Fominov2007} spin-orbit coupling\cite{Lv2011} or spin-dependent
potential,\cite{Terrade2013} up to date, the direct evidence of the spin-triplet supercurrent has
been observed in multilayer FM/FM/SC systems with a
non-collinear magnetization of the ferromagnetic layers.\cite{Klose2012,Gingrich2012,Leksin2012}

A first experimental hint for the long-range proximity effect was reported in a half-metal
Josephson junction based on CrO$_2$.\cite{Keizer2006} However, since the measured critical current
varied by two orders of magnitude in similar samples, the results of this experiment needed to be
confirmed. The strong evidence for the long range proximity effect was then reported
in the Josephson junctions based on Co.\cite{Khaire2010} The dependence of the critical current on
the Co layer thickness, which agrees with the theoretical expectations, provides a strong
experimental confirmation of the existence of the spin-triplet pairing in the FM/SC heterojunctions.
Further studies on the spin-triplet pairing concerned the FM/SC/FM and
FM/FM/SC junctions with a relative magnetization between the ferromagnets.
The spin triplet pairing in the clean FM/SC/FM nanostructures with an arbitrary angle between the 
magnetization of the FM layers  was theoretically studied by Haltermann et al. in
Refs.~\onlinecite{Halterman2005,Halterman2007,Halterman2008,Zhu2010}. The authors used the
self-consistent solutions of the microscopic Bogoliubov de-Gennes (BdG) equations and analyzed the
spin-triplet correlations as a function of the relative magnetization between the magnets. 
The self-consistent calculations allowed to confirm the experimentally observed angular dependence
of the critical temperature $T_c$ which monotonically increases due to the presence of
the long range spin-triplet correlations. $T_c$ reaches minimum if the relative
magnetization is parallel and maximum for antiparallel magnetization.\cite{Zhu2010} A different
behavior was observed for the FM/FM/SC nanostructures for which the critical
temperature is minimized in case of perpendicular alignment of the
magnetization.\cite{Leksin2011,Halterman2012}

Research on the spin-triplet pairing in the FM/FM/SC junctions has been
recently extended to systems with the conical (helical) ferromagnets (CM). Efforts to control the
long-range triplet supercurrent has been recently demonstrated in Josephson junctions based on
holmium(Ho)-Cobalt(Co)-holmium(Ho) multilayer setup.\cite{Robinson2010} One has been observed a
nonmonotonic dependence of the critical supercurrent as a function of the Ho layer thickness,
$d_{CM}$, with peaks for $d_{CM}=4.5$~nm and $10$~nm. By increasing the Co layer
thickness a slow decay of the critical current has been reported in agreement with
theoretical calculations.~\cite{Alidoust2010} Nevertheless, the theoretical model presented in
Ref.~\onlinecite{Alidoust2010} does not explain the complex dependence of the critical supercurrent
on the Ho thickness. The
nonmonotonic behavior of $I_C(d_{CM})$ has been obtained by Hal{\'a}sz et al. in
Ref.~\onlinecite{Halasz2011} who have performed calculations in the clean limit using Eliashberg
equations. Similar dependence has been also demonstrated by the use of
the Blonder-Tinkham-Klapwijk (BTK) approach.\cite{Wen2014,Jin2012}

In the mentioned theoretical works\cite{Halasz2011,Wen2014,Jin2012} the proximity effect
at the CM/SC interface has been neglected meaning that the superconducting
pair potential has been assumed to be a step function. However, as shown by recent
studies,\cite{Wu2014} only the self-consistent calculations of the tunneling conductance  guarantees
that the charge conservation law is satisfied. It means that one cannot
properly determine the tunneling conductance in the FM/CM/SC heterostructures by using the
non-self-consistent framework. The full self-consistent approach is needed.
The self-consistent calculations  of the spin-triplet correlations in
two layered CM/SC junctions have been presented in Refs.~\onlinecite{Wu2012,Wu2012PRL}.
Nevertheless, these studies concern only  the spin-triplet correlations between the CM
and SC. They do not include the analysis of the tunneling conductance (transport
calculations), the influence of the FM layer attached to the CM or the influence of the CM layer
thickness. Summing up, the theoretical analysis of the tunneling conductance through the FM/CM/SC
heterojunctions with the inclusion of the proximity effect in the full self-consistent framework has
not been presented until now.

In the present paper we report the full self-consistent calculations of the tunneling
conductance in the HM/CM/SC junctions. The charge transport in the considered system is mainly
determined by the anomalous Andreev reflection, the probability of which strongly depends on the
spin transmission in the CM layer. We consider the conductance in two cases in which the spin
transport is adiabatic and non-adiabatic. The conductance is analyzed over a broad range of
parameters determining the spiral magnetization in the CM. We show that the tunneling conductance
in the HM/CM/SC junctions strongly depends on the spin transport regime.
The paper is organized as follows: in Sec.~\ref{sec:theory} we introduce the basic concepts of the
theoretical scheme based on the self-consistent solution of the BdG equations and the BTK formalism.
In Sec.~\ref{sec:results} we present the results while the summary is included in
Sec.~\ref{sec:concl}.

\section{Theoretical method}
\label{sec:theory}
\begin{figure}[ht]
\begin{center}
\includegraphics[scale=.4]{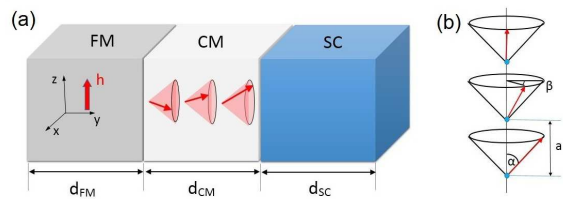} 
\end{center}
\caption{(a) (Color online) Schematic of the FM/CM/SC nanostructure. The exchange field
$\mathbf{h}$, denoted by the red arrow, is directed along the $z$ axis in the FM and has
a spiral structure in the CM. (b) The conical magnetic configuration with the
cone angle $\alpha$ and a rotational angle $\beta$. $a$ is the lattice constant. }
\label{fig1}
\end{figure}

We consider the FM/CM/SC structure schematically
illustrated in Fig.~\ref{fig1}. In the $x-z$ plane the system is assumed to
be infinite while the $y$ axis is perpendicular to the layers whose lengths are
denoted by $d_{FM}$, $d_{CM}$, $d_{SC}$, respectively. 
The value and the direction of the exchange field $\mathbf{h}$, denoted by the red arrow in
Fig.~\ref{fig1}, depends on the position. It is directed along the $z$ axis in the ferromagnet,
$\mathbf{h}=(0,0,h_{FM})$, while, in the conical ferromagnet, $\mathbf{h}$ is given by
\begin{equation}
 \mathbf{h}(y)= \left \{
 \begin{array}{l}
  h_{CM} \sin { \alpha }  \sin \left( \beta_0+\frac{\beta y}{a} \right ) \\
  h_{CM} \cos { \alpha }  \\
  h_{CM}\sin { \alpha }  \cos \left( \beta_0+\frac{\beta y}{a} \right ) \\
 \end{array} 
 \right . , 
 \label{eq:h}
\end{equation}
where $h_{CM}$ is the exchange field amplitude, $\beta_0$ determines the angle of a relative
magnetization of the CM layer measured from the FM layer at the FM/CM interface while $\alpha$ and
$\beta$ are the cone and rotation angle whose physical meaning is depicted in Fig.~\ref{fig1}(b).
From Eq. (\ref{eq:h}) the spatial period of the helix exchange field is $\lambda=2\pi
a/\beta$, where $a$ is the lattice constant. 

In the present paper we consider the charge transport through the FM/CM/SC junction and analyze
the tunneling conductance as a function of the system parameters.
As mentioned above, the correct analysis of the
conductance behavior require the inclusion of the proximity effect. This can be
done only by the full self-consistent calculations in which the pair potential
distribution is determined from the microscopic BdG equations.
The self-consistent procedure used in the paper consists of two steps. First, we determine
the self-consistent pair potential $\Delta(y)$ in the
nanostructure by solving BdG equations. Then, $\Delta(y)$ is used to calculate the tunneling
conductance within the BTK approach.~\cite{BTK,Bugoslavsky2005,Wojcik2011} Below,
both these steps are described in
detail.

\subsection{Self-consistent pair potential $\Delta(y)$ calculations}
\label{sec:model}
The effective BCS Hamiltonian of the considered system is given by
\begin{eqnarray}
\hat{\mathcal{H}}&=&\sum _{s} \int d^3 r
\hat{\Psi}^{\dagger}_s (\mathbf{r}) \left ( -\frac{\hbar ^2}{2m} \nabla ^2- \mu(\mathbf{r}) 
\right )
\hat{\Psi}_s(\mathbf{r}) \nonumber \\ 
&+& \frac{1}{2} \int d^3 r  \sum _{s s'} \left [ (i\sigma _y)_{ss'} \Delta
(\mathbf{r})\hat{\Psi}^{\dagger}_s(\mathbf{r})
\hat{\Psi}^{\dagger}_{s'}(\mathbf{r}) +H.c. \right ] \nonumber \\
&-& \int d^3 r  \sum _{s s'}  \hat{\Psi}^{\dagger}_s(\mathbf{r})
[\mathbf{h}(\mathbf{r}) \cdot \pmb{\sigma}]_{ss'} \hat{\Psi}^{\dagger}_{s'}(\mathbf{r}) \nonumber \\
&+&\int d^3r \frac{|\Delta(\mathbf{r})|^2}{g},
\label{eq:ham}
\end{eqnarray}
where $\hat{\Psi}^{\dagger}_s(\mathbf{r})$, $\hat{\Psi}_s(\mathbf{r})$ are the creation and
annihilation operators with spin $s$, 
$\mathbf{h}(\mathbf{r})=\big (h_x(\mathbf{r}),h_y(\mathbf{r}),h_z(\mathbf{r}) \big )$ is the
exchange field, $\pmb{\sigma}=(\sigma _x, \sigma _y, \sigma _z)$ is the Pauli matrices
vector, $\mu(\mathbf{r})$ is the chemical potential and $\Delta(\mathbf{r})$ is the spin-singlet
pair potential in real space defined as
\begin{equation}
 \Delta(\mathbf{r})=g(\mathbf{r}) \left < \hat{\Psi} _{\downarrow}(\mathbf{r})
\hat{\Psi}_{\uparrow}(\mathbf{r})  \right >, 
\label{eq:gap_def}
\end{equation}
where $g(\mathbf{r})$ is the phonon-mediated electron-electron coupling constant. 
The generalized Bogoliubov transformation 
\begin{equation}
 \hat{\Psi}_s(\mathbf{r})=\sum _n \left [ u_{ns}(\mathbf{r}) \gamma _n + \eta_s
v^*_{ns}(\mathbf{r}) \gamma^{\dagger}_n
\right ] ,
\end{equation}
where $\gamma _n$ and $\gamma^{\dagger}_n$ are the quasiparticle annihilation and creation
operators, $u_{ns}(\mathbf{r})$ and $v_{ns}(\mathbf{r})$ are the electron and hole components of
the amplitudes vector and $\eta_s= 1(-1)$ for spin down(up), reduces the Hamiltonian (\ref{eq:ham})
into the diagonal form. By the commutation relation 
\begin{equation}
[\hat{\Psi}_{\uparrow},\hat{\mathcal{H}}]=(H_e-h_z)\hat{\Psi}_{\uparrow}
-(h_x-ih_y)\hat{\Psi}_{\uparrow}+\Delta\hat{\Psi}_{\downarrow},
\end{equation}
\begin{equation}
[\hat{\Psi}_{\downarrow},\hat{\mathcal{H}}]=(H_e+h_z)\hat{\Psi}_{\downarrow}
-(h_x+ih_y)\hat{\Psi}_{\downarrow}+\Delta\hat{\Psi}_{\uparrow},
\end{equation}
where $H_e= -\frac{\hbar ^2}{2m} \nabla ^2- \mu(\mathbf{r}) $ is the single electron Hamiltonian,
and using the fact that the system is infinite in the $x-z$ plane we obtain the BdG equations in
the quasi-one dimensional form
\begin{widetext}
\begin{equation}
 \left (
 \begin{array}{cccc}
 H_e-h_z(y) & -h_x(y)+ih_y(y) & 0 & \Delta(y) \\
 -h_x(y)-ih_y(y) & H_e+h_z(y) & \Delta(y) & 0 \\
 0 & \Delta^*(y) & -H_e+h_z(y) & -h_x(y)-ih_y(y) \\
 \Delta^*(y) & 0 &  -h_x(y)+ih_y(y) & -H_e-h_z(y) \\
 \end{array}
 \right )  \\
 \left (
 \begin{array}{c}
 u_{n\uparrow}(y)\\
 u_{n\downarrow}(y)\\
 v_{n\uparrow}(y)\\
 v_{n\downarrow}(y)\\
 \end{array}
 \right )
 =
 E_n
 \left (
 \begin{array}{c}
 u_{n\uparrow}(y)\\
 u_{n\downarrow}(y)\\
 v_{n\uparrow}(y)\\
 v_{n\downarrow}(y)\\
 \end{array}
 \right ).
\label{eq:ham1D}
 \end{equation}
\end{widetext}
Equations (\ref{eq:ham1D}) are coupled with the expression for the pair potential given by
\begin{eqnarray}
 \Delta(y)=\frac{g(y)}{2} \sum _{|E_n|< \hbar \omega _D}  &\big [ & u_{n\uparrow}(y)
v^*_{n\downarrow}(y) + u_{n\downarrow}(y) v^*_{n\uparrow}(y) \big ] \nonumber \\
 &\times&  \left [ 1 - 2 f(E_n) \right ],
\label{eq:delta}
\end{eqnarray}
where $f(E)$ is the Fermi-Dirac distribution. The summation in
Eq.~(\ref{eq:delta}) is carried out only over the electronic states with energy $E_n$ inside the
Debye window $\left |E_n \right | < \hbar \omega _D$, where $\omega _D$ is the Debye frequency.
In our approach $g(\mathbf{r})$ is assumed to be nonzero only in the SC layer. \\

The self-consistent procedure used to solve the BdG equations (\ref{eq:ham1D}) is similar to these
reported in previous papers.\cite{Wu2012,Halterman2005,Halterman2007,Halterman2008} The main
difference is that the assumed basis functions have the form of the plane waves. Such choiceis
needed in the transport calculations because it guarantees nonzero current
at the boundaries of the system. The self-consistent procedure can be described as follows.
First, the BdG equations (\ref{eq:ham1D}) are diagonalized in the basis of the plane waves
\begin{equation}
  \left (
 \begin{array}{c}
 u_{n\uparrow}(y)\\
 u_{n\downarrow}(y)\\
 v_{n\uparrow}(y)\\
 v_{n\downarrow}(y)\\
 \end{array}
 \right )= \frac{1}{\sqrt{L}}
  \sum _q
 \left (
 \begin{array}{c}
 \tilde{u}_{nq\uparrow}\\
 \tilde{u}_{nq\downarrow}\\
 \tilde{v}_{nq\uparrow}\\
 \tilde{v}_{nq\downarrow}\\
 \end{array}
 \right ) \exp(ik_qy)
\end{equation}
where $\tilde{u}_{nq\uparrow},\tilde{u}_{nq\downarrow}, \tilde{v}_{nq\uparrow},
\tilde{v}_{nq\downarrow}$ are the expansion coefficients, $k_q=2\pi q/L $ is the wave vector  
with $q$ being an integer while $L$ is the total length of the nanostructure.
Then, using the calculated wave functions 
$(u_{n\uparrow},u_{n\downarrow},v_{n\uparrow}, v_{n\downarrow})^T$ the new pair potential
$\Delta(y)$ is determined on the basis of Eq.~(\ref{eq:delta}).
This new $\Delta(y)$ distribution is used in the next
iteration in which we again solve the BdG equations and determine $\Delta(y)$.
This procedure is repeated until the convergence is reached. 
Due to the high computational complexity of such scheme the parallel implementation of the numerical
procedure is required.

Finally, we also calculate the magnetization vector $\mathbf{m}$ given by the
formula
 \begin{eqnarray}
 m_x(y)&=&-\mu_{B} \sum _n \bigg \{
\big ( u_{n\uparrow}(y)u^*_{n\downarrow}(y)+u_{n\downarrow}(y)u^*_{n\uparrow}(y) \big )f(E_n)
\nonumber \\ &-& \big ( v_{n\uparrow}(y)v^*_{n\downarrow}(y)+v_{n\downarrow}(y)v^*_{n\uparrow}(y)
\big )[1-f(E_n)] \bigg \}, 
\label{eq:mx}
\end{eqnarray}
\begin{eqnarray}
m_y(y)&=&-i\mu_{B}\sum _n \bigg \{
\big ( u_{n\uparrow}(y)u^*_{n\downarrow}(y)-u_{n\downarrow}(y)u^*_{n\uparrow}(y) \big ) f(E_n)
\nonumber \\ &+& \big ( v_{n\uparrow}(y)v^*_{n\downarrow}(y)-v_{n\downarrow}(y)v^*_{n\uparrow}(y)
\big )
[1-f(E_n)] \bigg \}, 
\label{eq:my}
\end{eqnarray}
\begin{eqnarray}
m_z(y)&=&\mu_{B}\sum _n \bigg \{ \big (
u_{n\uparrow}(y)u^*_{n\uparrow}(y)-u_{n\downarrow}(y)u^*_{n\downarrow}(y) \big ) f(E_n)
\nonumber \\ 
&+& \big ( v_{n\uparrow}(y)v^*_{n\uparrow}(y)-v_{n\downarrow}(y)v^*_{n\downarrow}(y) \big )
[1-f(E_n)] \bigg \},
\label{eq:mz}
\end{eqnarray}
where $\mu_{B}$ are the Bohr magneton.

\subsection{Tunneling conductance calculations}
The tunneling conductance calculations have been performed within the tight-binding approximation
using the Kwant package.\cite{kwant} For this purpose we have transformed the BdG
equations (\ref{eq:ham1D}) into the discretized form on the grid $y_{\nu} = \nu a$
with lattice constant $a$ ($\nu = 1,2, \ldots$).
We introduce the discrete representation of the quasi-particle wave vector as follows:
$|\Psi(y_{\nu})\rangle
=
\left(|u^{\uparrow}(y_{\nu})\rangle,|u^{\downarrow}(y_{\nu})\rangle,|v^{\uparrow}(y_{\nu})\rangle,
|v^{\downarrow}(y_{\nu})\rangle \right)^T
\equiv |\Psi_{\nu}\rangle$,
Introducing a set $\pmb{\rho}$ of Pauli-like matrices in electron-hole space, the discretized
tight-binding form of the Hamiltonian in Eq.(\ref{eq:ham1D}) is given by
\begin{eqnarray}
 \label{HTB}
&&H= \sum _{\nu} \bigg \{ \rho _z \otimes \big [  (2t-\mu_{\nu})\mathbf{1}-h_{z\nu}\sigma_z \big ]
\bigg \} |\Psi_{\nu}\rangle \langle \Psi_{\nu} | \nonumber \\
&-& \sum _{\nu} \bigg \{ \rho _z \otimes \big ( t \mathbf{1} \big ) |\Psi_{\nu+1}\rangle \langle
\Psi_{\nu} | + H.c \bigg \} \\
&+& \sum _{\nu} \bigg \{ \big [ \big ( \Delta _{\nu} \rho _x - h_{x\nu}\mathbf{1} \big ) \otimes
\sigma _x \big ] - (h_{y\nu}\rho _z)\otimes \sigma _y \bigg \}
|\Psi_{\nu}\rangle \langle \Psi_{\nu} | \nonumber
\end{eqnarray}
where $\mu_{\nu}= \mu(y_{\nu})$, $(h_{x\nu},h_{y\nu},h_{z\nu})=\big (
h_{x}(y_{\nu}),h_{y}(y_{\nu}),h_{z}(y_{\nu}) \big )$,
$t=\hbar ^2/(2m a^2)$ and $\mathbf{1}$ is the unity matrix. 

Let us assume that the electron with
spin-up is injected from the FM into the SC through the CM layer. There are five possible
scattering processes: normal reflection with spin conservation $(R_{ee}^{\uparrow \uparrow})$,
normal reflection with spin-flip $(R_{ee}^{\downarrow \uparrow})$, reflection as a hole with
opposite spin (normal Andreev reflection, $R_{he}^{\downarrow \uparrow}$), reflection as a hole with
spin conservation (anomalous Andreev reflection, $R_{he}^{\uparrow \uparrow}$) and transmission as
a quasi-particle $T^{e\uparrow}$. In the above, $R_{e(h)e(h)}^{\uparrow(\downarrow)
\uparrow(\downarrow)}$ denotes the reflection probability where upper and lower right index
corresponds to the state of an incident particle while upper and lower left index is associated
with the reflected one, $T^{e\uparrow}$ is the transmission probability where upper indexes indicate
the state of incident particle. Analogous scattering processes can be distinguished for the
spin-down electron injected from the FM. Their probabilities are marked by $R_{ee}^{\downarrow
\downarrow}$, $R_{ee}^{\uparrow \downarrow}$, $R_{he}^{\uparrow \downarrow}$, $R_{he}^{\downarrow
\downarrow}$ and $T^{e\downarrow}$, respectively. \\
According to the BTK approach the current through the FM/CM/SC junction can be calculated from the
formula 
\begin{equation}
 I(V)=\int G(E) \left [ f(E-eV)- f(E) \right ] dE,
\end{equation}
where $V$ is the bias voltage and $f(E)$ is the Fermi-Dirac distribution. At low temperature the
energy dependent tunneling conductance $G(E)=\partial I / \partial V | _{V=E}$ (in units of $e^2
/h$) is given by
\begin{eqnarray}
G(E)&=&\frac{1+h_{FM}}{2} \bigg [ 1+R_{he}^{\uparrow \uparrow}(E)+
R_{he}^{\downarrow \uparrow}(E) \nonumber \\
& & \:\:\:\:\:\:\:\:\:\:\:\:\:\:\:\:\:\:\:\:\:\:\:\: - R_{ee}^{\uparrow
\uparrow}(E)-R_{he}^{\downarrow \uparrow}(E) \bigg ] \nonumber \\
&+& \frac{1-h_{FM}}{2}\bigg [ 1+ R_{he}^{\downarrow \downarrow}(E)+R_{he}^{\uparrow
\downarrow}(E) \nonumber \\
& & \:\:\:\:\:\:\:\:\:\:\:\:\:\:\:\:\:\:\:\:\:\:\:\: - R_{ee}^{\downarrow
\downarrow}(E)-R_{he}^{\uparrow \downarrow}(E), \bigg ]
\label{eq:g}
\end{eqnarray}
where $h_{FM}$, expressed in units of $\mu$, corresponds to the spin polarization at the Fermi
level in the FM layer.
In our calculations the reflection probabilities in Eq.~(\ref{eq:g}) are determined by the use of
the Kwant package\cite{kwant} which requires the implementation of the discretized tight-binding
Hamiltonian given by Eq.~(\ref{HTB}). In the paper, we consider the forward tunneling
conductance with the angle of the incident electron $\theta=0$ and neglect the scattering potential
at the interfaces.

\section{Results and discussion}
\label{sec:results}
In this section we analyze the tunneling conductance through the FM/CM/SC junctions by the use of
the full self-consistent approach presented in Sec.~\ref{sec:theory}. Since the first experimental
evidence for the long-range proximity effect (spin-triplet pairing) was reported in a half-metal
Josephson junction,\cite{Keizer2006} we restrict our analysis to the case in
which the ferromagnetic layer is embedded in a half-metal (HM), $h_{FM}=1$. 
In our calculations we neglect the Fermi wave-vector mismatch between the layers
assuming a constant value of the chemical potential $\mu$ throughout the nanostructure. Its value is
used as the energy unit. In the calculations we adopt the following values of the parameters: zero
temperature energy gap in the bulk $\Delta _0 = 0.01$, Debye energy $\hbar \omega _D=0.1$,
temperature $k_bT \approx 10^{-5}$ and the lattice constant $a=0.35$~nm corresponding to the conical
magnet Holmium.\cite{Robinson2010} 
Other parameters determining the magnetic configuration of the MC, namely the exchange
field amplitude $h_{CM}$, the cone angle $\alpha$, the rotation
angles $\beta$ and $\beta _0$, as well as the CM layer thickness $d_{CM}$ are used to analyze the
tunneling conductance through the HM/CM/SC junctions and vary from one simulation to another.

As predicted by Bergeret et al.\cite{Bergeret2001,Bergeret2003} the join effects of the Andreev
reflection and the proximity at the FM/SC interface allow for the coexistence of the spin-singlet
pairing correlations $ (|\uparrow \downarrow \rangle - |\uparrow \downarrow \rangle )/\sqrt{2}$ and
the spin-triplet pairing correlations with the total spin projection $m=0$ $(|\uparrow \downarrow
\rangle + |\uparrow \downarrow \rangle)/\sqrt{2}$). If a magnetically inhomogeneous layer, such as 
the CM layer, is present between the FM and SC, the spin-triplet state with $m=0$ can be rotated
to the state with $m=1$ ($|\uparrow\uparrow\rangle$). It means that the existence of the
spin-triplet pairing with $m=1$ strongly depends on the spin transmission in the CM layer. This, in
turn, is determined by the exchange field which in the CM has a rotating component varying with the
period $\lambda$. Depending on $\lambda$ the spins of electrons  injected from the FM can be
transmitted through the CM adiabatically - the spin orientation follows the spatial modulation of
the exchange field, or non-adiabatically - the period of the exchange field modulation is so short
that the electron spin is not able to adopt to the field changes. 
The degree of adiabaticity can be defined by the parameter $Q=\omega _L / \omega _h$, where 
$\omega_h=2\pi V_F / \lambda$ is the magnetic field modulation frequency in the electron's frame of
reference, $V_F$ is the Fermi velocity and $\omega _L=h_{CM}/ \hbar$
is the frequency of the spin Larmor precession.  In the adiabatic regime, $Q>>1$. 
Below, we analyze the tunneling conductance through the HM/CM/SC junctions in both adiabatic and
non-adiabatic regimes.

\subsection{Non-adiabatic regime}
\label{subsec:A}
All results presented in this subsection have been obtained for the rotational angle $\beta=30^o$
which corresponds to the spatial period of the helical exchange field $\lambda=3.4$~nm
measured in the Holmium.\cite{Robinson2010} For this value of $\lambda$ the spin transport through
the considered system is non-adiabatic,
$Q < 1$.

\begin{figure}[ht]
\begin{center}
\includegraphics[scale=.35]{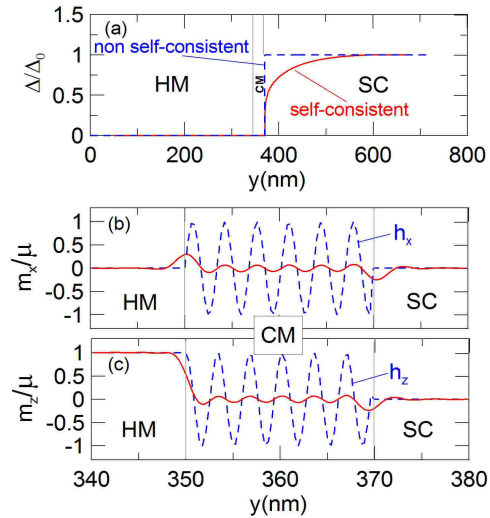} 
\end{center}
\caption{(Color online) (a) Comparison of the self-consistent (red solid line) and the
non-self-consistent (blue dashed line) pair potential $\Delta(y)$. (b) $m_x(y)$ and (c) $m_z(y)$
components of the magnetization in the nanostructure. For comparison the exchange field components
$h_x(y)$ and $h_z(y)$ are presented by the blue dashed lines. In panels (b) and (c) the vertical
gray dotted lines mark 
the boundaries of the CM layer.}
\label{fig2}
\end{figure}
In Fig.~\ref{fig2} we present the pair potential and the magnetization in the nanostructure 
calculated for $h_{CM}=1$, $\alpha=90^{\circ}$ and the thickness of the conical magnet layer
$d_{CM}=20$~nm.  
As shown in Fig.~\ref{fig2}(a) the self-consistent pair potential significantly differs from the one
used in the non-self-consistent approach. Due to the proximity effect $\Delta (y)$ does not have the
step-like form but smoothly increases in the SC region reaching
its bulk value $\Delta _0$ for a distance grater than the coherence length. 
Similarly, as the magnetism alters the superconductivity near the CM/SC interface, the
superconductivity also influences the magnetism. This so-called reverse proximity effect allows
to penetrate the magnetization into the SC region as presented in Fig.~\ref{fig2}(b,c). 
Almost a tenfold reduction of the magnetization amplitude in the CM layer (red
lines), as compared to the exchange field (blue dashed lines), results from the fact that the
chosen value of $\lambda$ corresponds to the non-adiabatic transport regime. In this regime the
changes of the exchange field seen by electrons flowing through the nanostructure are so fast that
the their spin do not have enough time to adopt to these changes. As a result, the  electron spin
rotates around the exchange field irregularly. Note
that, in accordance with Eqs.~(\ref{eq:mx})-(\ref{eq:mz}), the magnetization
is expressed as a sum of the averaged spin over states with different wave vectors. Since spins
of these states rotate around $\mathbf{h}$ with different irregular frequency, this sum averages to
low value. In subsection B we will show that the suppression of the magnetization does not dependent
on the conical magnet thickness $d_{CM}$ but, as expected, is mainly determined by the
spatial period of the helical exchange field $\lambda$. 
\begin{figure}[ht]
\begin{center}
\includegraphics[scale=.35]{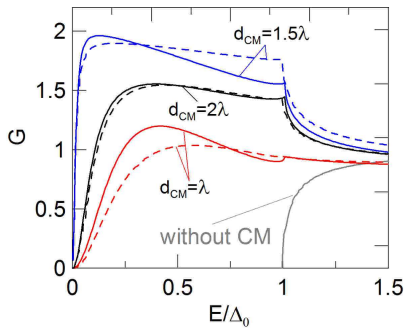} 
\end{center}
\caption{(Color online) Normalized conductance $G$ as a function of energy $E$
(in units of $\Delta _0$) for different thicknesses of the CM layer $d_{CM}=1,1.5$ and
$2$~$\lambda$. Results obtained by the use of the self-consistent (solid lines) and the
non-self-consistent (dashed lines) approach. The conductance
calculated without the CM layer is marked by the gray lines. Results for $h_{CM}=1$, $\alpha=90^o$
and $\beta _0=0$.}
\label{fig3}
\end{figure}

For the self-consistent pair potential $\Delta (y)$ we calculate the tunneling conductance
using the procedure described in Sec.~\ref{sec:theory} B. Figure~\ref{fig3} presents the normalized
conductance $G$ as a function of energy for different thicknesses of the CM
layer calculated by the use of the non-self-consistent (dashed lines) and the self-consistent (solid
lines) procedure. For comparison, we also mark the conductance calculated without the CM (gray
lines). As we see the dashed and solid gray lines overlap which results from the
fact that the conductance in this case is nonzero only for the high-energy limit, above the energy
gap, for which the electron incident into the superconductor does not experience much 
difference between the step-like pair potential and the smooth pair potential from the
self-consistent approach. Results presented in Fig.~\ref{fig3} clearly show that the self-consistent
conductance is considerably different than this obtained in the non-self-consistent framework. The
most pronounced difference between them is observed in the subgap energy range. Based on
the results presented in Fig.~\ref{fig2} and \ref{fig3} one can formulate the following conclusion:
\textit{ to properly determine the tunneling conductance in the HM/CM/SC heterojunctions the full
self-consistent calculations including the proximity effect are needed.}

\begin{figure}[ht]
\begin{center}
\includegraphics[scale=.35]{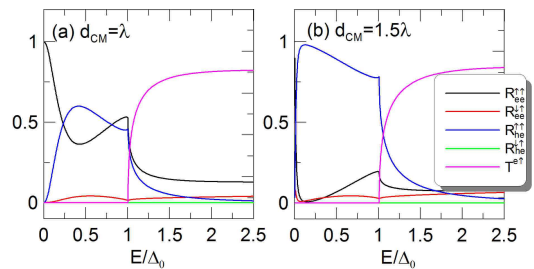} 
\end{center}
\caption{(Color online) The reflection and transmission probabilities $R_{ee}^{\uparrow \uparrow}$,
$R_{ee}^{\downarrow \uparrow}$, $R_{he}^{\downarrow \uparrow}$, $R_{he}^{\uparrow \uparrow}$,
$T^{e\uparrow}$ as a function of energy (in units of $\Delta _0$) for the CM layer thicknesses
(a) $d_{CM}=\lambda$ and (b) $d_{CM}=1.5~\lambda$ for which the conductance in the
subgap region reaches the minimum and maximum value, respectively (see Fig.~\ref{fig3}).
Results for $h_{CM}=1$, $\alpha=90^o$ and $\beta _0=0$.
}
\label{fig4}
\end{figure}

To explain the conductance behavior presented in Fig.~\ref{fig3} let us first discuss the
half-metal/superconductor (HM/SC) junctions without the CM layer. In the HM/SC junctions the normal
Andreev reflections are forbidden due to the isolation of the spin band. This leads to the
zero conductance in the subgap energy regime as depicted by the gray solid lines in Fig.~\ref{fig3}
(we assume no scattering potential at the interface). The situation diametrically changes if we
put the conical magnet between the HM and SC. As predicted by Bergeret et
al.\cite{Bergeret2001,Bergeret2003} the magnetic inhomogeneity at the FM/SC interface can induce
the non-zero correlations of all three components $m=0$ and $m=\pm 1$ of the spin-triplet state.
As a consequence, one appears an extra scattering mechanism, called
the anomalous Andreev reflection in which electron incident into the SC is
reflected as a hole with the same spin (in contrast to the normal Andreev reflection in which the
incident electron and the reflected hole have opposite spins). 
For the HM/CM/SC junctions, this new scattering mechanism, if it exists, leads to the nonzero
conductance in the subgap region as presented in Fig.~\ref{fig3}. As one can see the conductance
strongly depends on the thickness of the conical magnet layer, $d_{CM}$, i.e. its value reaches
minimum for $d_{CM}=\lambda$ and maximum for $d_{CM}=1.5\lambda$, respectively. In Fig.~\ref{fig4}
we present the reflection and transmission probabilities $(R_{ee}^{\uparrow \uparrow})$,
$(R_{ee}^{\downarrow \uparrow})$, $R_{he}^{\downarrow \uparrow}$, $R_{he}^{\uparrow \uparrow}$,
$T^{e\uparrow}$ as a function of energy for these two distinguished thicknesses.
We see that for the CM thickness $d_{CM}=1.5\lambda$ the increase
of the conductance in the subgap region is mainly determined by the increase of the anomalous
Andreev reflection probability $R_{he}^{\uparrow \uparrow}$. On the other hand the probability
$R_{he}^{\uparrow \uparrow}$ is suppressed for $d_{CM}=\lambda$ for which the normal reflection
with spin conservation $(R_{ee}^{\uparrow \uparrow})$ emerges and also contributes to the
conductance value leading to its decrease. Regardless of the CM layer thickness, for low energy,
the anomalous Andreev
reflection probability $R_{he}^{\uparrow \uparrow}$ drops to zero while the normal reflection
$R_{ee}^{\uparrow \uparrow}$ increases to unity. This results in the zero conductance at zero
energy as demonstrated in Fig.~\ref{fig3}.

\begin{figure}[ht]
\begin{center}
\includegraphics[scale=.3]{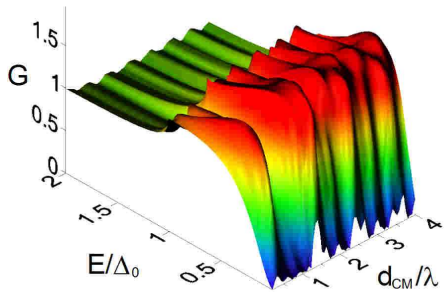} 
\end{center}
\caption{(Color online) Normalized tunneling conductance $G$ as a function of energy $E$
(in units of $\Delta_0$) and CM layer thickness $d_{CM}$ (in units of $\lambda$). Results for
$h_{CM}=1$, $\alpha=90^o$ and $\beta _0=0$. }
\label{fig5}
\end{figure}

\begin{figure}[ht]
\begin{center}
\includegraphics[scale=.35]{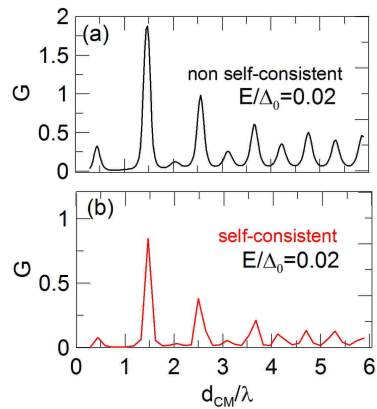} 
\end{center}
\caption{(Color online) Normalized tunneling conductance $G$ as a function of
CM layer thickness $d_{CM}$ (in units of $\lambda$) for energy $E/\Delta_0=0.02$. Results obtained
by the use of (a) the non-self-consistent and (b) the self-consistent approach. Results for
$h_{CM}=1$, $\alpha=90^o$ and $\beta _0=0$.}
\label{fig6}
\end{figure}

Now, we discuss the thickness dependence of the conductance $G(d_{CM})$, important from
the viewpoint of experiments in which the critical current is measured as a function of the CM layer
thickness.  In Fig.~\ref{fig5} we present the tunneling conductance as a function of
energy and $d_{CM}$. 
We see that the conductance oscillates as a function of the CM layer thickness with the
amplitude which is not constant but varies wit $d_{CM}$.
The clear evidence of these oscillations is shown in Fig.~\ref{fig6}(b) where the cross-section of
the $G(E,d_{CM})$ map is presented for the energy $E/\Delta _0=0.02$.
The small number of points in this figure results from the high computational cost
of the full-self consistent calculations. 
Results presented in Fig.~\ref{fig6}~(b) are consistent with the phenomena observed in experiments,
namely (i) the conductance is a nonmonotonic function of $d_{CM}$ with peaks for the half-integer
multiplies of $\lambda$ and (ii) the conductance slowly decays with increasing CM layer
thickness. For comparison, in Fig.~\ref{fig6}(a) the non-self-consistent dependence $G(d_{CM})$
for $E/ \Delta _0=0.02$ are also presented. As we see the peaks in conductance calculated in the
non-self-consistent framework are greater than the corresponding peaks calculated self-consistently.
Moreover, the conductance decay rate (with increasing $d_{CM}$) is slower than in the
self-consistent approach.

The irregular nonmonotonic dependence of the conductance $G(d_{CM})$ presented in Figs.~\ref{fig5}
and \ref{fig6} can be explained as follows.
If the spin-active region (the CM layer) is present at the FM/SC interface the anomalous
Andreev reflections can appear giving raise to the nonzero conductance in the subgap
region.~\cite{Bergeret2001,Bergeret2003} The probability $R_{he}^{\uparrow \uparrow}$ depends
mainly in the spin transition in the CM layer. Note, that the strength
of the spin-flip scattering in the CM is proportional to the off-diagonal matrix elements of the
Hamiltonian (\ref{eq:ham1D}) which have the form $-h_x(y)+ih_z(y)$. Since $h_x(y)$ and $h_z(y)$ 
vary periodically, the strength of the spin-flip scattering also oscillates with increasing CM
layer thickness. Nevertheless in the non-adiabatic regime the modification of the exchange field
seen by electrons flowing through the nanostructure is so fast that the electronic spins are not
able to follow these changes. It entails the irregular oscillations of the anomalous
Andreev reflection probability and, in consequence, the irregular oscillations of the conductance
depicted in Figs.~\ref{fig5} and \ref{fig6}.

\begin{figure}[ht]
\begin{center}
\includegraphics[scale=.45]{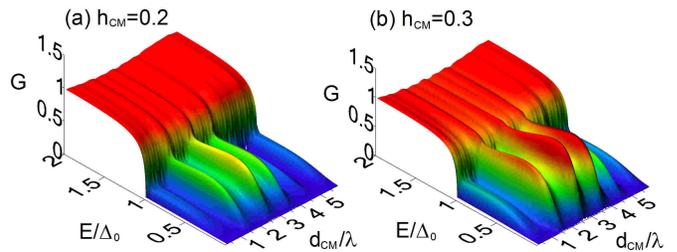} 
\end{center}
\caption{(Color online) Normalized tunneling conductance $G$  as a function of energy
$E$ (in units of $\Delta_0$) and CM layer thickness $d_{CM}$ (in units of $\lambda$) for the
amplitude of the exchange field in the CM layer (a) $h_{CM}=0.2$ and (b) $h_{CM}=0.3$. Other
parameters are the same as in previous calculations.}
\label{fig7}
\end{figure}

\begin{figure}[ht]
\begin{center}
\includegraphics[scale=.3]{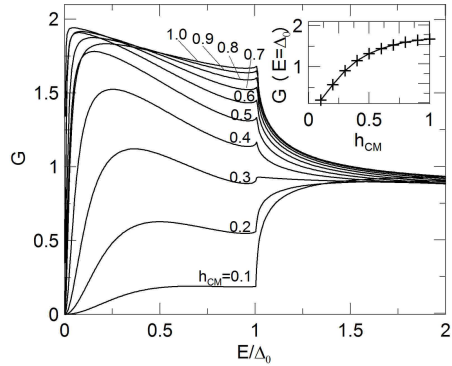} 
\end{center}
\caption{(Color online) Normalized tunneling conductance $G$ as a function of
energy $E$ (in units of $\Delta_0$) for different values of the exchange field amplitude $h_{CM}$.
Results for $d_{CM}=2.5\lambda$.}
\label{fig8}
\end{figure}

In delineating the role of the spin-triplet pairing in the charge transport through the HM/CM/SC
junctions, it is necessary to understand the behavior of the conductance under the influence of the
magnetic configuration in the CM layer determined by the value of the exchange field
amplitude $h_{CM}$ and the angles $\alpha$ and $\beta _0$ (see Eq.~\ref{eq:h}). In
Fig.~\ref{fig7} we plot the conductance maps $G(E,d_{CM})$ for the exchange field amplitude (a)
$h_{CM}=0.2$ and (b) $h_{CM}=0.3$. Other parameters are assumed to be the same as in previous
calculations, results of which are presented in Fig.~\ref{fig5}.
Figure \ref{fig7} clearly demonstrates the suppression of the conductance in the subgap region with
decreasing exchange field amplitude $h_{CM}$. The comparison of Fig.~\ref{fig7}
and Fig.~\ref{fig5} allows to conclude that the strength of this conductance suppression depends on
the CM layer thickness. It is minimal for $d_{CM}=2.5\lambda$. Note that, for $h_{CM}=0.2$ the
conductance peak for $d_{CM}=2.5\lambda$ is still well pronounced [Fig.~\ref{fig7}(a)]. Further
reduction of the amplitude $h_{CM}$ leads to the situation in which the conductance peak survives
only for $d_{CM}=2.5\lambda$ in consistency with the experimental measurements reporting the peak of
the critical current exactly for this value of the CM layer thickness. Although the conductance for
$d_{CM}=2.5\lambda$ decreases slower than for other thicknesses, even for this value of
$d_{CM}$ the conductance in the subgap region is suppressed with decreasing $h_{CM}$. This
suppression is depicted in Fig.~\ref{fig8} which presents $G(E)$ for different values of
the exchange field amplitude $h_{CM}$. We see that in the limit $h_{CM} \rightarrow 0$, as expected,
the conductance in the subgap region tends to zero.
We should also notice clear cups in the conductance for the energy $E=\Delta _0$. The non-self
consistent analysis for the FM/SC junctions without the CM layer leads to the analytical
dependence of $G(E=\Delta _0,h_{FM})$ which is a decreasing function of the exchange
field in the ferromagnet\cite{Zutic2000}
\begin{equation}
G(E=\Delta_0,h_{FM})=\frac{4\sqrt{1-h_{FM}^2}}{1+\sqrt{1-h_{FM}^2}}. 
\end{equation}
As depicted in the insert of Fig.~\ref{fig8}
for the HM/CM/SC junctions the dependence $G(E=\Delta_0,h_{CM})$ is an increasing function of
$h_{CM}$ in contrast to the FM/SC structure. This results from
the fact that the conductance for the considered energy is mainly determined by the anomalous
Andreev reflections whose probability increases with increasing amplitude of the spiral magnetic
configuration in the CM layer.

\begin{figure}[ht]
\begin{center}
\includegraphics[scale=.35]{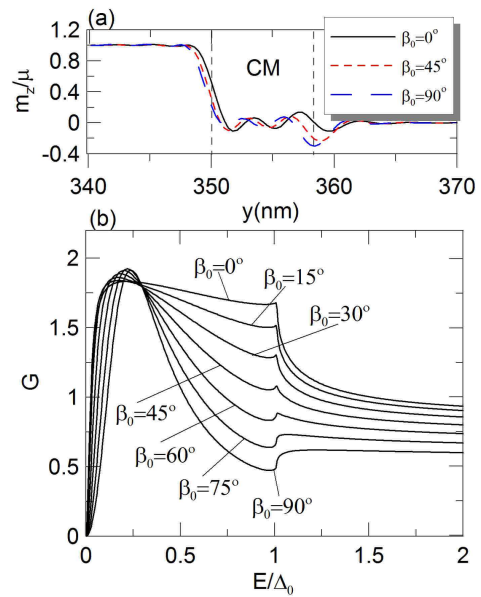} 
\end{center}
\caption{(Color online) (a) $z$-component of the magnetization $m_z(z)$ for three different angles
$\beta _0$ of a relative magnetization of the CM layer measured from the HM layer at the HM/CM
interface. Dotted vertical lines marked boundaries of the CM layer.  (b) Normalized
tunneling conductance $G$  as a function of energy $E$ (in units of $\Delta _0$) for different
angles $\beta _0$. Results for $d_{CM}=2.5\lambda$, $\alpha=90^o$ and $\beta _0=0$.}
\label{fig9}
\end{figure}

The magnetic configuration in the CM layer can be modified not only by changing the amplitude
$h_{CM}$ but also by changing the spatial configuration determined by the angles $\alpha$, $\beta$
and $\beta _0$. Figure \ref{fig9}(b) presents the conductance $G(E)$ for different angles $\beta
_0$ of a relative magnetization of the CM layer measured from the HM layer at the HM/CM interface.
In panel (a) we present the $z$-component of the magnetization $m_z(z)$ for three different angles
$\beta _0$. Note that $m_z(y)$ is not discontinuous but changes smoothly at the interfaces HM/CM
and CM/SC. It saturates to unity in the HM region on the left-hand and penetrates the SC region on
the right-hand. Although the amplitude of the exchange field $\mathbf{h}$ is the same for all three
cases, the conductance in the subgap region decreases with increasing $\beta_0$ [see
Fig.~\ref{fig9}(b)]. This behavior can be easily understood by considering two factors. The
fist is directly related to the oscillatory dependence of the
conductance with the CM layer thickness. In fact, the introduction of a relative
magnetization between the HM and CM layers corresponds to the
phase shift in the oscillatory dependence  of the helical magnetic configuration. Therefore, in the
first approximation, the dependence $G(d_{CM})$  should
be shifted in argument by $\Delta d_{CM}=2\pi a/\beta _0$. This shift causes the conductance for
$d_{CM}=2.5\lambda$ (corresponds to the maximum value for $\beta _0 =0$) shifts to
lower value related to  $G(2.5\lambda + \Delta d)$. The second factor is the increase of the
normal reflection probability at the HM/CM interface resulting from the discontinuity of the
exchange field. The presented dependence $G(\beta _0)$ is important in the simulation of the real
HM/CM/SC structure since the conical magnets used in the
multilayer setup have several ways to orient the magnetic moments with respect to the half-metal
magnetization depending on the magnetic coupling at the HM/CM interface.

\begin{figure}[ht]
\begin{center}
\includegraphics[scale=.45]{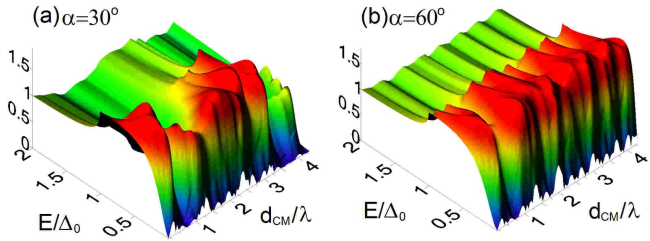} 
\end{center}
\caption{(Color online) Normalized tunneling conductance $G$ as a function of energy $E$
(in units of $\Delta_0$) and CM layer thickness $d_{CM}$ (in units of $\lambda$) calculated for (a)
$\alpha=30^o$ and (b) $\alpha=60^o$. Results for $h_{CM}=1$, $\beta _0=0$.}
\label{fig10}
\end{figure}

\begin{figure}[ht]
\begin{center}
\includegraphics[scale=.45]{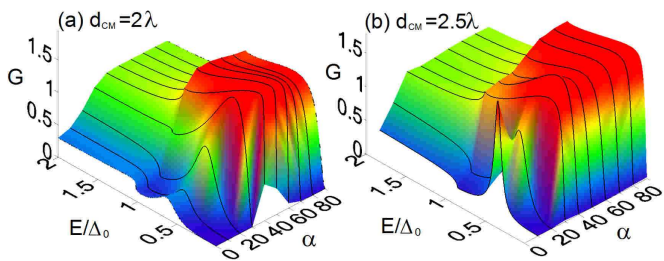} 
\end{center}
\caption{(Color online) Normalized tunneling conductance G as a function of
energy $E$ (in units of $\Delta_0$) and the cone angle $\alpha$ for the CM layer thickness (a)
$d=2\lambda$ and (b) $d=2.5\lambda$. Results for $h_{CM}=1$, $\beta _0=0$.}
\label{fig11}
\end{figure}

All results presented so far have been obtained for $\alpha=90^o$ for which the $y$-component
of the exchange field $h_y$ is zero - the magnetization in the CM layer rotates in the $x-z$ plane.
Now, we analyze the conductance for the non-zero value of $h_y$. In Fig.~\ref{fig10}
we plot the conductance map $G(E,d_{CM})$ for two values of the cone angle (a) $\alpha=30^o$ and (b)
$\alpha=60^o$. We see that for the low cone angle $\alpha=30^o$ the conductance map
$G(E,d_{CM})$ significantly differs from the map for $\alpha=90^o$ (Fig.~\ref{fig5}).
Based on Fig.~\ref{fig10}  one can state that regardless of the angle $\alpha$
the value of the conductance in subgap region oscillates as a function of $d_{CM}$ whereas
the period and the amplitude of these oscillations are irregular. In Fig.~\ref{fig11} we present the
conductance map $G(E,\alpha)$ as a function of the cone angle calculated  for the CM layer thickness
(a) $d=2\lambda$ and (b) $d=2.5\lambda$~nm.
As we can see, the conductance in the subgap region decays with decreasing the cone angle, whereas
the decay rate strongly depends on $d_{CM}$. For $d=2\alpha$ it is stronger
than for $d=2.5\alpha$.

\subsection{Adiabatic regime}
In this subsection we analyze the the tunneling conductance in the HM/CM/SC
junctions in the adiabatic regime for a long period $\lambda$.
In subsection \ref{subsec:A} we have demonstrated that in the non-adiabatic regime the magnetization
in the CM layer is strongly suppressed compared to the helical exchange field. As presented in
Fig.~\ref{fig2}~(b,c) the amplitude of $m_x(y)$ and $m_z(y)$ modulation in the CM  is
about ten times smaller than the amplitude of the helical exchange field.
Such a strong suppression have been obtained for a short period of the exchange field modulation
$\lambda=3.4$~nm corresponding to $Q<1$ (non-adiabatic regime). It has been suggested that the
magnitude of this suppression can be used as the additional parameter to measure the degree of
adiabaticity. 

\begin{figure}[ht]
\begin{center}
\includegraphics[scale=.35]{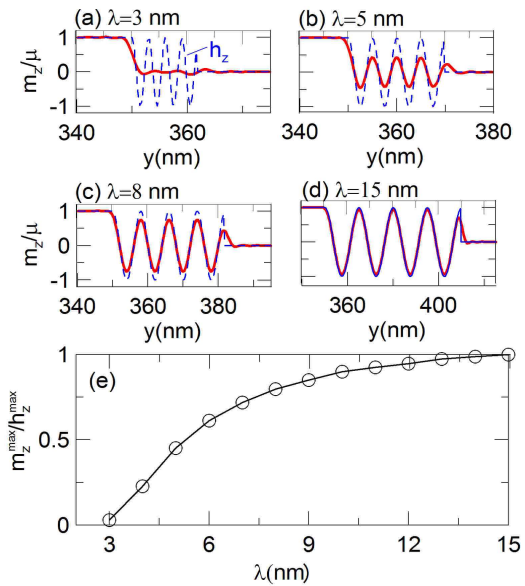} 
\end{center}
\caption{(Color online) $z$-component of the magnetization (red lines) for
different periods of the exchange field modulation (a) $\lambda=3$~nm, (b) $\lambda=5$~nm, (c)
$\lambda=8$~nm, (d) $\lambda=15$~nm. The distributions of the exchange field are plotted by the blue
dashed lines. (e) Ratio of the amplitudes $m_z^{max}/h_z^{max}$ in the CM as a function of
the period of the exchange field modulation $\lambda$. Results for $h_{CM}=1$, $\alpha=90^o$,
$\beta _0=0^o$ and $d_{CM}=4 \lambda $.}
\label{fig12}
\end{figure}
\begin{figure*}[ht]
\begin{center}
\includegraphics[scale=0.8]{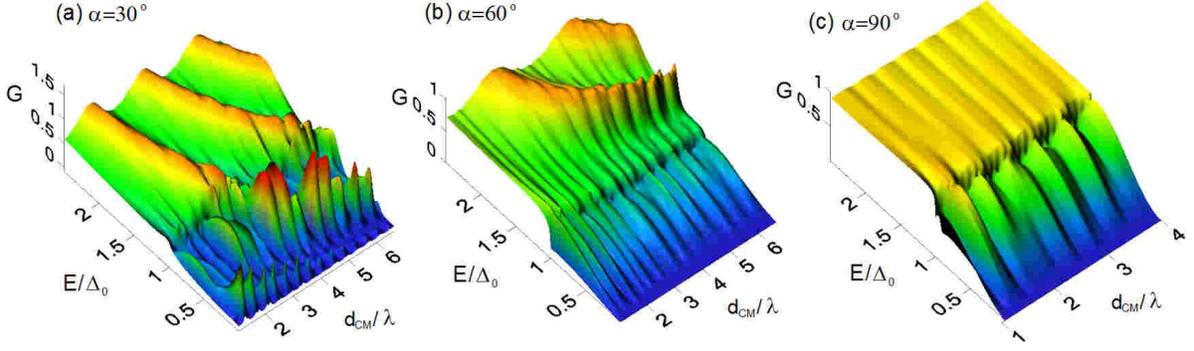} 
\end{center}
\caption{(Color online) Normalized tunneling conductance $G$ as a function of energy $E$
(in units of $\Delta_0$) and CM layer thickness $d_{CM}$ (in units of $\lambda$) calculated for (a)
$\alpha=30^o$, (b) $\alpha=60^o$ and (c) $\alpha=90^o$. Results for $\lambda=15$~nm, $h_{CM}=1$
and $\beta _0=0$.}
\label{fig13}
\end{figure*}
In Fig.~\ref{fig12} we demonstrate the self-consistent $z$-component of the magnetization for
different values of $\lambda$ assuming $h_{CM}=1$. For comparison the distributions of the exchange
field are plotted by the blue dashed lines.
As presented in Fig.~\ref{fig12} (compare the red and blue lines), the suppression of the
magnetization in the CM layer is more pronounced for a short period of the exchange field modulation
$\lambda$ (non-adiabatic regime) and almost completely disappears for a long $\lambda$ (adiabatic
regime). For $\lambda=15$~nm there is no difference between the magnetization $m_z(y)$ and
the exchange field $h_z(y)$ except the boundaries of the CM layer where $m_z(y)$ smoothly changes
penetrating the SC region due to the reverse proximity effect. Fig.~\ref{fig12}~(e) presents the
ratio of the amplitudes $m_z^{max}/h_z^{max}$ in the CM as a function of
the exchange field modulation period $\lambda$. As one can see the ratio saturates to
the value $m_z^{max}/h_z^{max}=1$ for $\lambda$ grater than $15$~nm for which the transport through
the heterojunction can be assumed to be adiabatic.
The further analysis presented in this subsection will be carried out in the adiabatic regime
for $\lambda=15$~nm corresponding to $Q\approx 10$.

Figure~\ref{fig13} shows the normalized conductance as a function of energy and CM layer thickness
for the cone angles $\alpha=30^o,60^o$ and $90^o$.
For $\alpha=90^o$ corresponding to $h_y=0$ [Fig.\ref{fig13}(c)] the conductance in
the subgap region oscillates with a period $\lambda/2$. It reaches maximum for $d_{CM}$ being an
integer multiple of $\lambda/2$. In this case (we assume $\beta _0=0$) the spin of electrons
injected from the HM has the same direction as the exchange field in the CM.
In the adiabatic regime the spin of electrons flowing through the nanostructure follows
the exchange field. Therefore, in this case the anomalous Andreev reflection probability is exactly
proportional to the off-diagonal elements $-h_x(y)+ih_y(y)$ which oscillate leading to the regular
oscillations of the conductance presented in Fig.~\ref{fig13}(c). This behavior considerably
differs from the irregular conductance oscillations presented in Fig~\ref{fig5} for non-adiabatic
regime. Moreover, note, that the value of
the conductance in the subgap region is lower than this obtained in the non-adiabatic regime
(compare with Fig.~\ref{fig5}) which is caused by the enhancement of the normal reflection
probability due to increase of the exchange field modulation period $\lambda$.

\begin{figure}[ht]
\begin{center}
\includegraphics[scale=0.38]{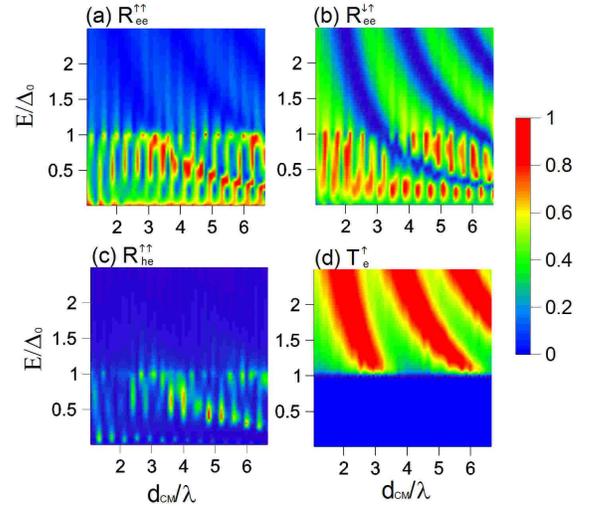} 
\end{center}
\caption{(Color online) The reflection and transmission probabilities (a)
$R_{ee}^{\uparrow\uparrow}$, (b) $R_{ee}^{\downarrow\uparrow}$, (c) $R_{he}^{\uparrow\uparrow}$
and (d) $T^{\uparrow}_e$ as a function of energy $E$
(in units of $\Delta_0$) and CM layer thickness $d_{CM}$ (in units of $\lambda$) for
$\alpha=30^o$. The normal Andreev reflection probability $R_{he}^{\downarrow\uparrow}(E,d_{CM})=0$.
 Results for $\lambda=15$~nm, $h_{CM}=1$ and $\beta _0=0$.}
\label{fig14}
\end{figure}

For the cone angle $\alpha \neq 90^o$, corresponding to the non-zero $h_y$, the spin of electrons
injected from the HM is non-collinear with the exchange field at the HM/CM interface.
Therefore, the electronic spin starts to precesses around the exchange field direction with the
Larmor frequency which, in the adiabatic regime, is much higher than the frequency of the exchange
field modulation experienced by electrons flowing through the CM.  For $\lambda=15$~nm
($Q\approx10$) one period of the exchange field modulation corresponds to ten full-rotations
 of electron spin around the exchange field direction. Therefore, for the non-zero $h_y$ the spin
behavior in the CM layer is determined by joint effects: Larmor precession and the exchange field
modulation. 
For certain  energies and the CM thicknesses this complex spin behavior leads to the
enhancement of normal reflection probability with spin flip $R_{ee}^{\downarrow \uparrow}$
presented in Fig.~\ref{fig14}(b). Note, that the probability of this scattering mechanism in the
non-adiabatic regime is close to zero (Fig.~\ref{fig4}).
\begin{figure}[ht]
\begin{center}
\includegraphics[scale=0.35]{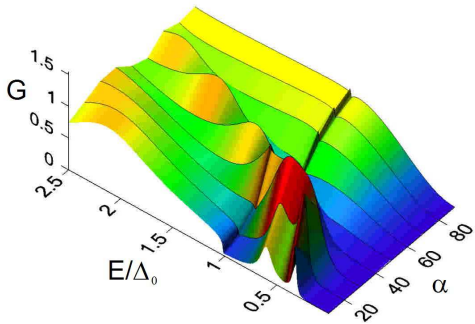} 
\end{center}
\caption{(Color online) Normalized tunneling conductance $G$ as a function of energy $E$
(in units of $\Delta_0$) and the cone angle $\alpha$. Results for $\lambda=15$~nm and
$d_{CM}=4\lambda.$}
\label{fig15}
\end{figure}
According to Eq.~\ref{eq:g}, in the range in which $R_{ee}^{\downarrow
\uparrow}$ increases, the conductance is suppressed leading to the characteristics $G(E,d_{CM})$
demonstrated in Fig.~\ref{fig13}(a). In this figure the conductance suppression for the energy above
$\Delta _0$ corresponds to green areas. Nevertheless this suppression expands also on the subgap
region for which $R_{ee}^{\downarrow \uparrow}$ is even grater than above $\Delta _0$
[Fig.~\ref{fig14}(b)]. The evolution of the conductance with increasing cone angle $\alpha$ is
presented in Fig.~\ref{fig15}.
We see that the conductance in the subgap region initially decreases and then clear conductance peak
for $\alpha=30^o$ appears. The position of this peak is shifted on the energy scale with
increasing  cone angle $\alpha$.

\begin{figure}[ht]
\begin{center}
\includegraphics[scale=0.35]{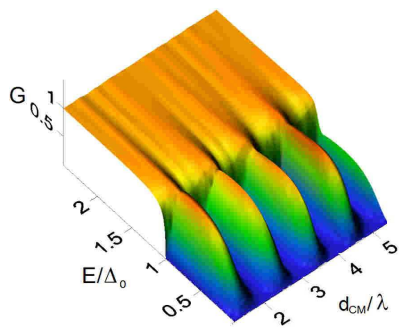} 
\end{center}
\caption{(Color online) Normalized tunneling conductance $G$ as a function of energy $E$
(in units of $\Delta_0$) and CM layer thickness (in units of $\Delta _0$). Results for
$\alpha=90^o$ and $h_{CM}=0.2$.}
\label{fig16}
\end{figure}

Finally, in Fig.~\ref{fig16} we present the conductance map $G(E,d_{CM})$ calculated for
$\alpha=90^o$ and the exchange field amplitude $d_{CM}=0.2$. As shown in previous subsection in the
non-adiabatic regime the decrease of $h_{CM}$ results in decrease of the conductance in subgap
region. This conductance suppression is different for different CM
thicknesses (see Fig.~\ref{fig7}) leading to the conductance peak for $d_{CM}=2.5\lambda$ in
consistency with the experimental results.~\cite{Robinson2010}
In the adiabatic regime the exchange field amplitude $h_{CM}$ affects the conductance in a different
manner. It leads to the conductance  decay for the thicknesses being an integer multiples of
$\lambda$ whereas for $d_{CM}=(N/2)\lambda$ the conductance remains almost unchanged. Therefore, the
conductance in the sugap region oscillates regularly with the period $\lambda$ as presented in
Fig.~\ref{fig16}.

\section{Summary}
\label{sec:concl}
We present the detailed analysis of the transport properties in the HM/CM/SC junction within 
the fully self-consistent framework based on the combined BTK formalism and the BdG equations. For
comparison, the calculations have been also carried out with the use of the non-self consistent
scheme. One has been shown that the peaks in the CM layer thickness dependence of the
conductance are significantly reduced when the self-consistent procedure is applied (cf. Fig.
\ref{fig6}). It is clear from our analysis that to properly determine the tunneling conductance in
the HM/CM/SC heterojunctions the full self-consistent calculations including the proximity effect
should be carried out.

Due to the spin band separation in the HM, the anomalous Andreev reflection mechanism which appears
in structures with magnetically inhomogeneous layer (such as the CM layer) results in the nonzero
conductance within the subgap region (cf. Fig. \ref{fig3}). Its probability 
strongly depends on the spin transmission in the CM layer. Therefore, we analyze the influence of
the exchange field modulation $\lambda$ on the behavior of spins of electrons transfered through the
CM layer. With this respect, we show that one can distinguish between two regimes. In the 
non-adiabatic regime (low value of the exchange field modulation $\lambda$) the changes 
of the exchange field are so fast that the spins of
electrons are not able to adopt and, as a result, they rotate around $\mathbf{h}$ irregularly and 
as an average give a small value of magnetization in comparison to the amplitude of the exchange
field. On the other hand, in the adiabatic regime (high value of $\lambda$) the magnetization coming
from the spins of electrons is almost identical to the exchange field position dependence within the
CM layer (cf. Fig \ref{fig12}). 
The conductance behavior as function of the energy and the CM layer thickness has different behavior
in the two mentioned regimes (cf Figs. \ref{fig5} and \ref{fig13}(c)). The regular oscillations
observed in the adiabatic regime are caused by the fact that anomalous Andreev reflection
probability is exactly proportional to the off-diagonal elements of the
Hamiltonian $-h_x(y)+ih_y(y)$. This leads to constant hight of the peaks in the CM layer thickness
conductance dependence, whereas for the case of the non-adiabatic regime the hight of those peaks is
decreasing with increasing thickness, $d_{CM}$, in consistency with the experiment observation (cf.
Fig. \ref{fig6}). Moreover in the non-adiabatic regime the decrease of the exchange filed amplitude
results in the well pronounced conductance peak for $d_{CM}=2.5\lambda$ for which the peak of the
critical current was observed in the experiment.\cite{Robinson2010} 
The influence of other parameters characterizing the exchange field behavior 
in the CM layer are also analyzed (such as the $\alpha$, $\beta$ and $\beta_0$ angles). It is shown
that the conductance is strongly affected by the geometrical structure  of the exchange
field determined by the cone angle $\alpha$ and the rotational a%merlin.mbs apsrev4-1.bst 2010-07-25
4.21a (PWD, AO, DPC) hacked
ngle $\beta$ and $\beta _0$.

\section*{Acknowledgments}
This work  was financed from the budget for Polish Science in the years 2013-2015. Project number:
IP2012 048572.

%\bibliography{refs}

\end{document}